\documentclass[aps,twocolumn,amsmath,amssymb,superscriptaddress,prb]{revtex4-1}

\usepackage{graphicx}
\usepackage{dcolumn}
\usepackage{subfigure}
\usepackage{bm}
\usepackage{color}

\begin{document}


\title{Magnetism in Co$_{1-{\rm x}}$Fe$_{\rm x}$Sb$_{3}$ skutterudites from density functional theory}

\author{Mikael R{\aa}sander}\email{m.rasander@imperial.ac.uk}
\affiliation{%
Department of Materials and Nanophysics, KTH Royal Institute of Technology, Electrum 229, SE-164 40 Kista, Sweden
}%
\affiliation{%
Department of Materials, Imperial College London, Exhibition Road, London SW7 2AZ, United Kingdom
}%
\author{Lars Bergqvist}
\affiliation{%
Department of Materials and Nanophysics, KTH Royal Institute of Technology, Electrum 229, SE-164 40 Kista, Sweden
}%
\affiliation{%
SeRC (Swedish e-Science Research Centre), KTH, SE-100 44 Stockholm, Sweden}%

\date{\today}

\begin{abstract}
We have investigated the electronic and magnetic properties of Co$_{1-{\rm x}}$Fe$_{\rm x}$Sb$_{3}$ skutterudites from density functional theory and Monte Carlo simulations. We find that above a certain threshold in the Fe concentration, somewhere between x=0.125 and x=0.25, Co$_{1-{\rm x}}$Fe$_{\rm x}$Sb$_{3}$ is ferromagnetic with an atomic moment which increases asymptotically towards about 1~$\mu_{B}$/Fe and a non-zero Curie temperature which reaches 70~K for FeSb$_{3}$. Ferromagnetism is favored due to a Stoner instability in the electronic structure, where a large density of states at the Fermi-level makes it favorable to form the ferromagnetic ground state. 
\end{abstract}

\maketitle
\section{introduction}\label{sec:intro}
Many materials have been proposed for high efficiency energy conversion in thermoelectric devices, e.g. the clathrates and zintl phases.\cite{Snyder2008} The skutterudites, generally represented in the form R$_{\rm y}$M$_{4}$X$_{12}$, where M is a transition metal, e.g. Co or Ir, X is a pnictogen, e.g. As or Sb, and R, if present, is typically a rare-earth element, e.g. La, have shown promise for high thermoelectric conversion efficiency. However, the efficiency for energy conversion, described by the figure-of-merit $ZT=S^2\sigma T/\kappa$, where $S$ is the Seebeck coefficient, $\sigma$ is the electric condictivity, $\kappa$ is the thermal conductivity, and $T$ is temperature, is not optimal. The main drawback in the skutterudites, exemplified for instance by CoSb$_{3}$, is that the thermal conductivity is found to be too high.\cite{Hammerscmidt2013} In order to improve the figure-of-merit of skutterudites, it is possible to include heavy filler elements in the MX$_{3}$ skutterudite framework, as in the case of LaFe$_{4}$Sb$_{12}$, or to alloy the skutterudite phase, as in the case of Co$_{1-{\rm x}}$Fe$_{\rm x}$Sb$_{3}$.\cite{Yang2000,Katsuyama1998,Amornpitoksuk2007}
\par
Co$_{1-{\rm x}}$Fe$_{\rm x}$Sb$_{3}$ has received attention since doping of CoSb$_{3}$ leads to more favourable thermoelectric properties, e.g. a larger carrier concentration and lowered thermal conductivity.\cite{Yang2000,Katsuyama1998,Amornpitoksuk2007} We note that studies on Co$_{1-{\rm x}}$Fe$_{\rm x}$Sb$_{3}$ contain rather small amounts of Fe; the maximum Fe content has been varied in these studies from ${\rm x}=0.1$\cite{Yang2000} to 0.4\cite{Katsuyama1998} mostly since larger amounts of Fe is not possible to incorporate into the skutterudite phase with traditional means, due to the metastability of FeSb$_{3}$ which separates into FeSb$_{2}$ and Sb.\cite{Hornbostel1997} Recently, however, attention has been given to pure FeSb$_{3}$ since it has been shown that it is possible to synthesize thin FeSb$_{3}$ films using the modulated elemental reactant method (MERM).\cite{Hornbostel1997,Mochel2011,Daniel2015}
\par
Following the synthesis of FeSb$_{3}$ films, density functional calculations have revealed that the pure FeSb$_{3}$ system is a ferromagnet with a low transition temperature;\cite{Rasander} much lower than room temperature and also much lower than the thermoelectric operating temperatures suggested for these compounds. FeSb$_{3}$ was also found to be softer than CoSb$_{3}$, for both the elastic constants and the lattice dynamics, which suggests that the thermal conductivity in this material is lower than in CoSb$_{3}$.\cite{Mochel2011,Rasander}  It was also shown by R{\aa}sander {\it et al.} that magnetism is essential in order to have a dynamically stable system.\cite{Rasander} Interestingly, Fe has been found to be paramagnetic in Co$_{1-{\rm x}}$Fe$_{\rm x}$Sb$_{3}$,\cite{Yang2000} as well as in other antimonide skutterudites.\cite{Danebrock1996,Sales1997,Gajewski1998,Yang2000} However, there are ferromagnetic skutterudites found in the literature, e.g. NaFe$_{4}$Sb$_{12}$ and KFe$_{4}$Sb$_{12}$,\cite{Leithe-Jasper2003,Leithe-Jasper2004} and ferromagnetic behavior should therefore not be surprising. 
\par
Experimentally, FeSb$_{3}$ was found to be paramagnetic with an effective paramagnetic moment of 0.57(6)~$\mu_{B}$/f.u..\cite{Mochel2011}  However, density functional theory calculations do obtain a ferromagnetic solution for the ground of FeSb$_{3}$.\cite{Rasander,Xing2015,Daniel2015} In fact, FeSb$_{3}$ is found to be a near half-metal in its ferromagnetic ground state\cite{Rasander} while experiments performed by M{\"o}chel  {\it et al.}\cite{Mochel2011} obtain a small gap of 16.3(4)~meV. This behavior agrees with a recent hybrid density functional study of Lemal {\it et al.}\cite{Lemal2015} which finds a small gap of 33~meV. In their study Lemal {\it et al.} also determine the FeSb$_{3}$ phase to be antiferromagnetic with an estimated N{\'e}el temperature of 6~K and an atomic magnetic moment of 1.6~$\mu_{B}$/Fe.  The latter should be compared with previously reported ferromagnetic and antiferromagnetic moments of 1.0~$\mu_{B}$/Fe and 1.1~$\mu_{B}$/Fe, respectively.\cite{Rasander} Note, however, that the experiment performed by Daniel {\it et al.}\cite{Daniel2015} find the FeSb$_{3}$ films to have a metallic conductivity and therefore no gap, so the experimental situation regarding FeSb$_{3}$ is not completely clear. What is clear, however, is that magnetism plays an important role in FeSb$_{3}$ and, therefore, it should also be of importance in Co$_{1-{\rm x}}$Fe$_{\rm x}$Sb$_{3}$. In order to reconcile the discrepancy between experimental and theoretical results regarding the magnetic properties, it was suggested by Xing {\it et al.}\cite{Xing2015} that a possible overestimation of the magnetic properties compared to experiments could possibly be due to spin fluctuations associated with a nearby quantum critical point that are strong enough to renormalize the mean-field magnetic state predicted by density functional calculations,\cite{Xing2015} similar to what is observed in ferropnictides.\cite{Mazin2004} Such behavior has indeed been observed in NaFe$_{4}$Sb$_{12}$ and KFe$_{4}$Sb$_{12}$ where experimentally ordered moments of $\sim1$~$\mu_{B}$ is found compared to $\sim3$~$\mu_{B}$ predicted by theory.\cite{Leithe-Jasper2003,Leithe-Jasper2004}
\par
In this paper, we provide results from density functional calculations on Co$_{1-{\rm x}}$Fe$_{\rm x}$Sb$_{3}$ in order to provide a further understanding of the magnetic properties of these systems. Specifically, we have calculated the transition temperatures based on a first principles approach. We show that, among other things, the magnetic moments are to a great extent localised on the Fe atoms and the transition temperatures are low throughout the whole composition range in Co$_{1-{\rm x}}$Fe$_{\rm x}$Sb$_{3}$.
\par
The paper is outlined as follows: In section~\ref{sec:method} we present the details of the calculations, in section~\ref{sec:results} we present the results of our study and finally in section~\ref{sec:conclusions} we summarise our finding and draw conclusions. 

\section{computational details}\label{sec:method}
\begin{figure}[t]
\includegraphics[width=8cm]{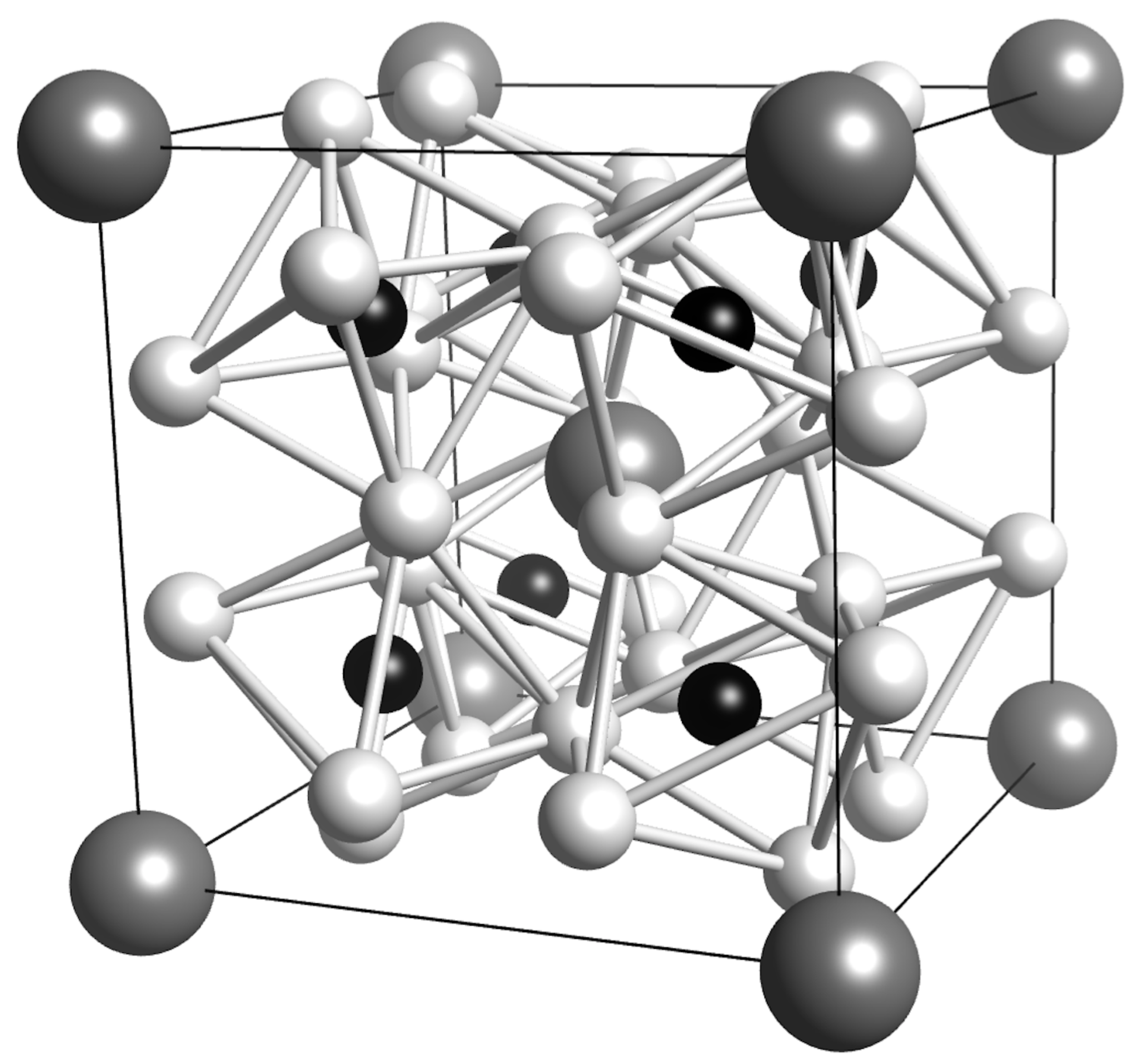}
\caption{\label{fig:structure} Illustration of the crystal structure of the skutterudite structure. Fe (black spheres) is residing inside canted octahedral cages of Sb (white spheres). Filler atoms are presented by large grey spheres.}
\end{figure}
The binary skutterudite structure has a unit cell containing four formula units with body centered cubic lattice vectors and belongs to the space group Im$\bar{3}$ (No. 204), where metal atoms and pnictogen atoms occupy the 8$c$ and 24$g$ positions respectively. The pnictogen atoms occupy the general position (0,$y$,$z$). The skutterudite framework, i.e. MX$_{3}$, contains large voids at the 2$a$ positions of the lattice. In Fig.~\ref{fig:structure} we show the conventional unit cell of the skutterudite structure (8 formula units) with the voids, at (0,0,0) and ($\frac{1}{2}$,$\frac{1}{2}$,$\frac{1}{2}$), filled with a rare-earth element. 
\par
In order to calculate the magnetic properties of Co$_{1-{\rm x}}$Fe$_{\rm x}$Sb$_{3}$, we have made use of a combination of first principles density functional calculations. Structural properties of CoSb$_{3}$ and FeSb$_{3}$ have been determined using the Projector augmented wave (PAW) method\cite{Blochl} as it is implemented in the Vienna ab initio simulation package (VASP).\cite{KresseandFurth,KresseandJoubert} The Perdew-Burke-Ernzerhof (PBE) generalised gradient approximation was used for the exchange-correlation energy functional. Further details regarding these calculations can be found in Ref.~\onlinecite{Rasander}. The resulting structural parameters are shown in Table~\ref{tab:fesb3}. As can be seen in Table~\ref{tab:fesb3} FeSb$_{3}$ has a larger lattice constant than CoSb$_{3}$. However, the $y$ and $z$ parameters for the Sb atoms are very similar in FeSb$_{3}$ and CoSb$_{3}$. We have therefore chosen to apply a Vegard's law behaviour for the intermediate phases in Co$_{1-{\rm x}}$Fe$_{\rm x}$Sb$_{3}$, for $0<{\rm x}<1$, with the $y$ and $z$ parameters fixed. A Vegard's law type of behaviour is also supported by Yang {\it et al.}\cite{Yang2000} in their study of Co$_{1-{\rm x}}$Fe$_{\rm x}$Sb$_{3}$, for $0\leq{\rm x}\leq0.1$, where such a behaviour was observed. The lattice constant for FeSb$_{3}$ labelled Vegard's law in Table~\ref{tab:fesb3} is extrapolated from the result of Yang {\it et al.}\cite{Yang2000}

\begin{table}[t]
\caption{\label{tab:fesb3} Comparison of the evaluated lattice constants and crystallographic $y$ and $z$ values for the Sb atoms in FeSb$_{3}$ and CoSb$_{3}$ for spin-polarized (in ferromagnetic (FM) and anti-ferromagnetic (AFM) configurations) and non spin-polarized (NSP) calculations. The Vegard's law value is extracted from the study on Co$_{1-{\rm x}}$Fe$_{\rm x}$Sb$_{3}$ by Yang {\it et al.}\cite{Yang2000}}
\begin{ruledtabular}
\begin{tabular}{ccccc}
System &  &  $a$ (\AA) & $y$ & $z$ \\ 
 \hline
FeSb$_{3}$ & NSP & 9.153 & 0.327  & 0.160 \\
FeSb$_{3}$ &  FM & 9.178& 0.331 & 0.160 \\
FeSb$_{3}$ & AFM & 9.166  & 0.331 & 0.159 \\
FeSb$_{3}$ & Expt.\cite{Hornbostel1997} & 9.176 & 0.340 & 0.162 \\
FeSb$_{3}$ & Expt. ($T=10$~K)\cite{Mochel2011} & 9.212 & 0.340 & 0.158 \\
FeSb$_{3}$ & Expt. ($T=300$~K)\cite{Mochel2011} & 9.238 & 0.340 & 0.157 \\
FeSb$_{3}$ & Vegard's law\cite{Yang2000} & 9.126 & - & - \\
\\
CoSb$_{3}$ & NSP & 9.115 & 0.333  & 0.160 \\
CoSb$_{3}$  & Expt.\cite{Schmidt1987} & 9.039 & 0.335 & 0.158 \\
 \end{tabular}
\end{ruledtabular}
\end{table}

\par
In addition to the previously mentioned PAW calculations, the electronic and magnetic properties of Co$_{1-{\rm x}}$Fe$_{\rm x}$Sb$_{3}$ has been investigated using the Korringa-Kohn-Rostoker (KKR) method using the SPR-KKR software package\cite{SPRKKR}. The electronic states were treated fully relativistic by solving the Dirac equation employing a basis set consisting of $s$, $p$ and $d$ orbitals. The local spin density approximation (LSDA) was used for the exchange correlation functional and the shape of the potential was treated using the atomic sphere approximation (ASA) with additional empty spheres to minimize the overlap of the different muffin-tin spheres centered on the atomic sites. The coherent potential approximation (CPA)\cite{soven}, was employed to treat disorder effects, i.e mixing of Co and Fe. Magnetic moments and exchange parameters within a Heisenberg model were obtained from the magnetic force theorem using the Lichtenstein-Katsnelson-Antropov-Gubanov (LKAG) formula\cite{sasha1,sasha2} and then used as input for subsequent Monte Carlo simulations using the UppASD package\cite{UppASD}. Our accuracy in determining the transition temperatures is estimated to give an error of about $\pm3$~K.

\begin{figure}[t]
\includegraphics[width=9cm]{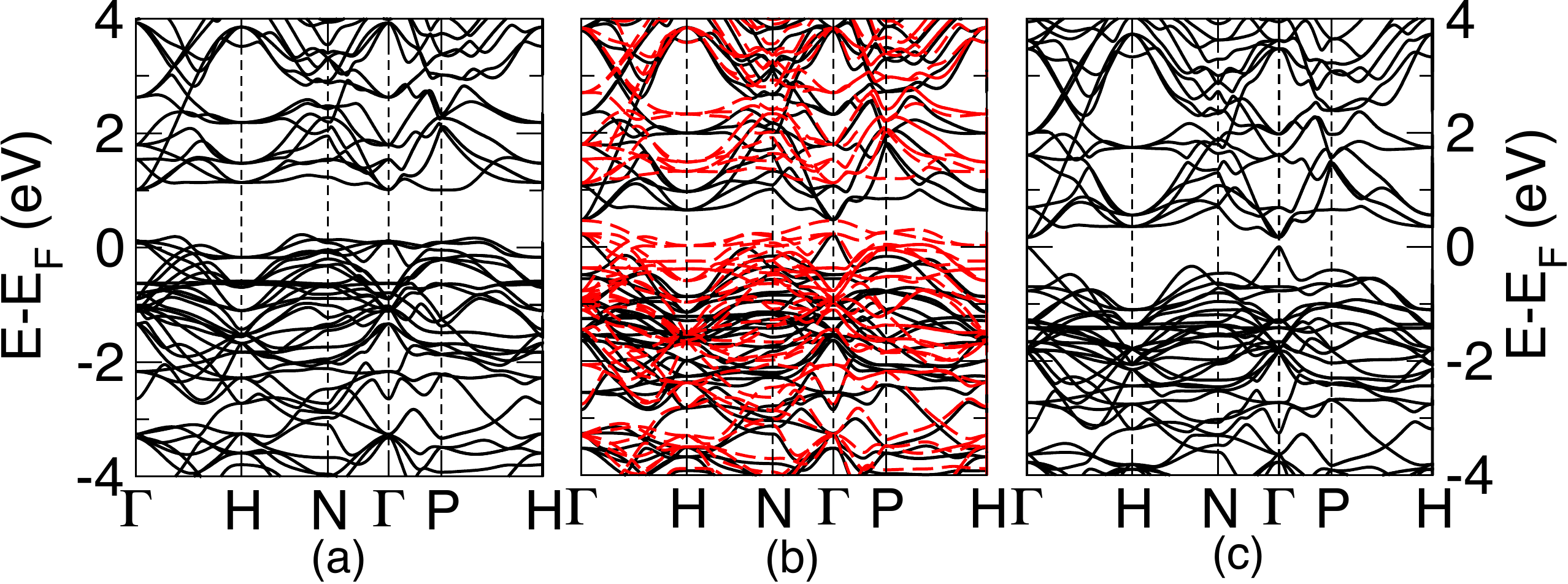}
\caption{\label{fig:bands} (Color online) Calculated band structures of (a) NSP FeSb$_{3}$, (b) FM FeSb$_{3}$ (c) and CoSb$_{3}$. The Fermi-level, $E_{F}$, is found at 0 eV. In (b) the spin-up band are plotted with solid (black) lines and the spin-down bands are plotted with dashed (red) lines. For more details, see Ref.~\onlinecite{Rasander}.}
\end{figure}
\section{Results}\label{sec:results}

We begin with a discussion of the electronic structure of the two endpoint systems in our study, namely CoSb$_{3}$ and FeSb$_{3}$. The electronic structure of these compounds have been discussed in detail before\cite{Rasander,Xing2015,Daniel2015} and we will here briefly recapture the results of previous studies. The band structures in CoSb$_{3}$ and FeSb$_{3}$ are shown in Fig.~\ref{fig:bands}. CoSb$_{3}$ is a semiconductor with a small direct gap of 0.17~eV. FeSb$_{3}$ on the other hand is a near half-metal ferromagnet (FM) where the spin-up channel is almost full. Non-spin polarised (NSP) FeSb$_{3}$ has a large number of bands with rather small dispersion close to the Fermi-level which according to the Stoner criterion makes it favourable for the system to have a ferromagnetic ground state.\cite{Rasander} The magnetic moment in FM FeSb$_{3}$ is $\sim1.0$~$\mu_{B}$/Fe\cite{Rasander,Xing2015} largely localised to the Fe atoms. 
\begin{figure}[b]
\includegraphics[width=5cm,angle=270]{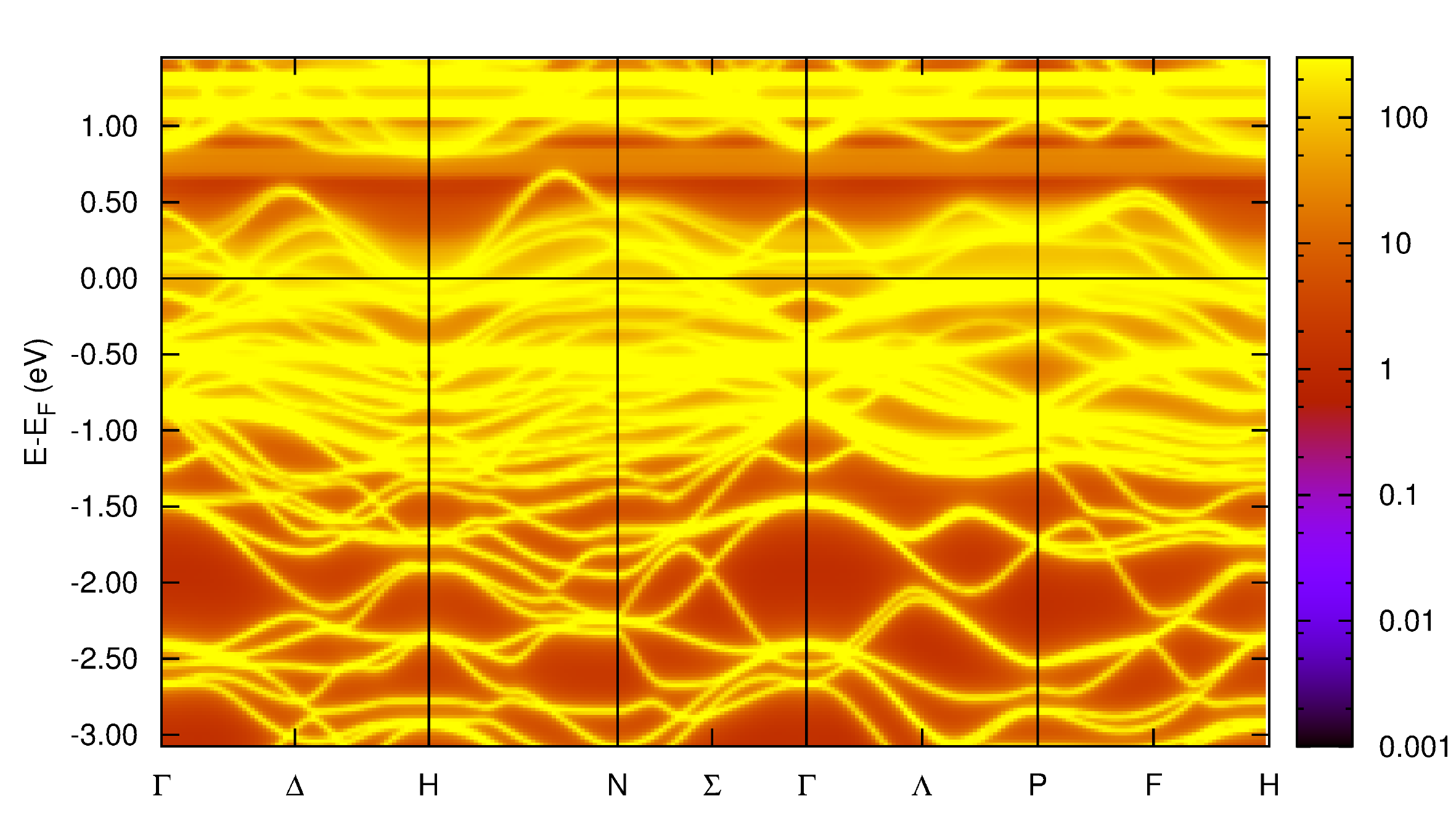}
\caption{\label{fig:BSF} Calculated Bloch spectral function (BSF) for Co$_{0.75}$Fe$_{0.25}$Sb$_{3}$ (spin polarized calculation) along directions of high symmetry in the Brillouin zone. The y-axis represents the energy in eV and the Fermi-level, $E_{F}$ is found at 0 eV.}
\end{figure}

\par
In Fig.~\ref{fig:BSF} we show the calculated Bloch spectral function (BSF) $A(E,\mathbf{k})$ for Co$_{0.75}$Fe$_{0.25}$Sb$_{3}$ within CPA. The BSF is a suitable way of analyzing band structures in disordered systems and can be seen as a wave vector $\mathbf{k}$-resolved density of states (DOS) function. For ordered systems the $A(E,\mathbf{k})$ is a $\delta$-like function at energy $E$ and can be used as an alternative way to calculate band structures, as in Fig.~\ref{fig:bands}. However, for disordered systems, in addition to the energy, the BSF also has a broadening associated to the amount of disorder in the system. From Fig.~\ref{fig:BSF}, it is clear that disorder mostly takes place close to Fermi level since the electron bands are most diffuse there and several bands are crossing the Fermi level. The underlying mechanism for this are explained in the following way. Adding Fe to CoSb$_{3}$, $d$-bands from Co and Fe start to cross the Fermi level and the density of states increases making it favourable to form a ferromagnetic ground state and this effect becomes stronger as the Fe content increases. 

\begin{figure}[t]
\includegraphics[width=8cm,angle=0]{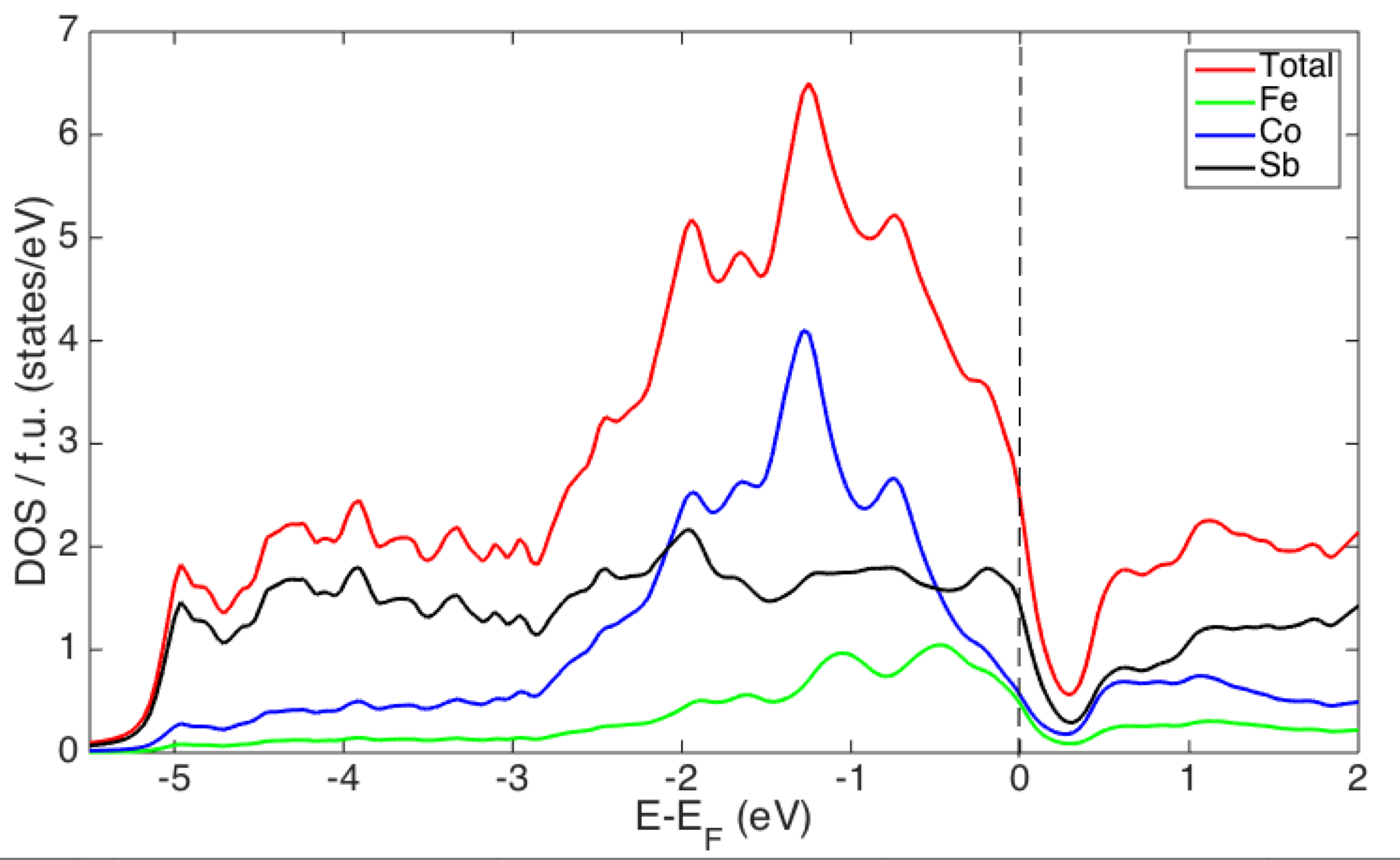}
\caption{\label{fig:DOS} Calculated NSP density of states (DOS) / formula unit (f.u) for Co$_{0.75}$Fe$_{0.25}$Sb$_{3}$. The Fermi level, $E_F$, is marked with a vertical dashed line. }  
\end{figure}

This is analyzed further by calculating the NSP density of states of Co$_{0.75}$Fe$_{0.25}$Sb$_{3}$, shown in Fig.~\ref{fig:DOS}. According to the Stoner criteria of ferromagnetism, a material may lower its total energy by exchange splitting of the bands, i.e. become ferromagnetic, if $I \cdot D(E_F)> 1$, where $D(E_F)$ is the NSP density of states at the Fermi level, $E_{F}$, and $I$ is the Stoner exchange integral. $I$ has been calculated for most elements in the periodic table and its value is typically around 0.7-0.8 eV\cite{janak} without much variation. If the Stoner criteria is assumed to hold, it means that $D(E_F)$ must exceed $\approx$ 1.3 states/eV in order to promote ferromagnetism. From Fig.~\ref{fig:DOS}, taking only into account Fe and Co states, $D(E_F)$ is around 1.1 states/eV which is on the borderline to become ferromagnetic. If also Sb states are taken into account, or partially through hybridization with Fe and Co states, then the Stoner criteria is fulfilled and ferromagnetism is expected. The analysis is somewhat simplified but it indicates that doping with Fe of CoSb$_{3}$ promotes a magnetic state, which is indeed also found in the full electronic structure calculations.

\begin{figure*}[t]
\includegraphics[width=14cm]{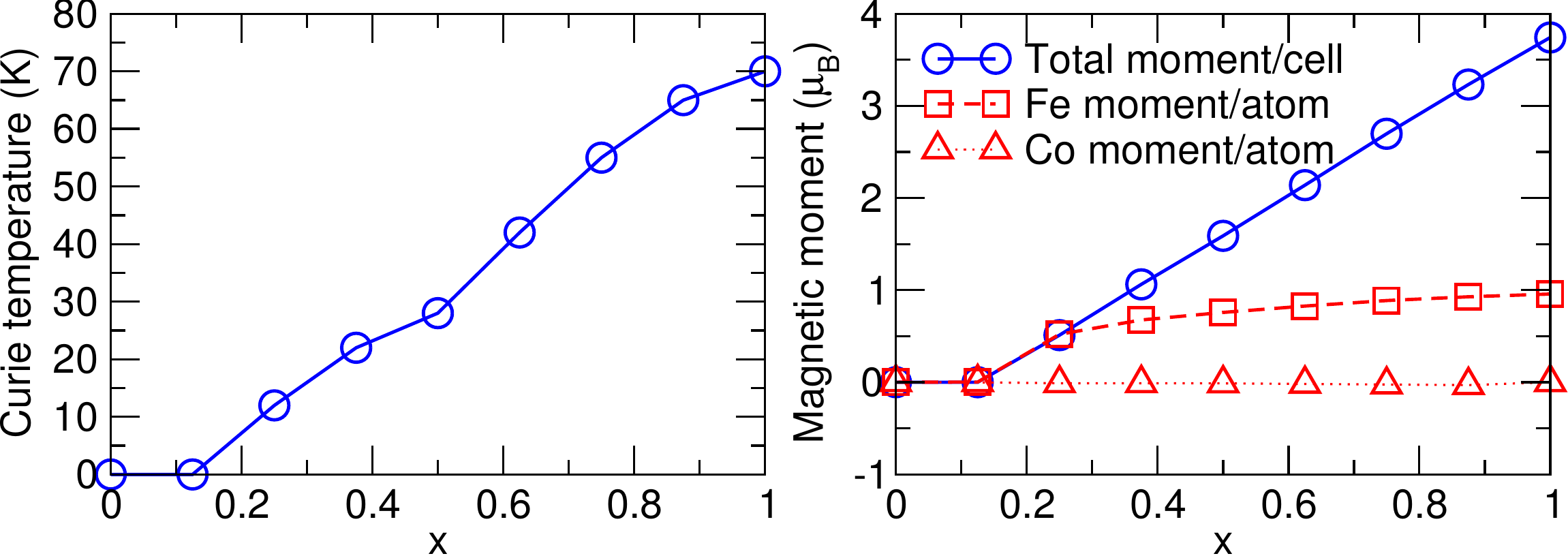}
\caption{\label{fig:data} Calculated transition temperatures (left) and magnetic moments (right) for Co$_{1-{\rm x}}$Fe$_{{\rm x}}$Sb$_{3}$. Note that a summation of the Fe moment in the case of x~$=1$ will give a moment that is larger than the total moment shown in the figure.}
\end{figure*}
\par
In Fig.~\ref{fig:data} and Table~\ref{tab:data} we show the calculated magnetic moments and transition temperatures for $0\leq{\rm x}\leq1$. We note that the formation of a magnetic moment sets in for ${\rm x}>0.125$. After this point we find a linear increase in the total magnetic moment with increasing Fe concentration. The moments localised on the Fe atoms increase drastically for $0.125\leq{\rm x}\leq0.25$ and thereafter the increase is much less dramatic as the Fe content increases. The Fe moment has its maximum value of 0.96~$\mu_{B}$ for ${\rm x}=1$, i.e. for FeSb$_{3}$. We note that the Fe moments make up most of the total moments. The moments localised on the other atoms work towards lowering the total moment in the system. This is represented by the very small negative Co moments shown in Fig.~\ref{fig:data} and Table~\ref{tab:data}.
\par
In order to investigate the build up of magnetism for small Fe concentrations further, we have performed calculations on a single Fe atom substituting Co in a 3$\times$3$\times$3 CoSb$_{3}$ unit cell using the PAW method. The effective concentration of Fe in CoSb$_{3}$ is in this case less than 1\%.  We find that for this low concentration the total magnetic moment is zero and the moments on the different atoms are also zero which is in agreement with our ASA calculations. This suggests that a rather substantial concentration of Fe is necessary for the formation of a magnetic moment in the system. 
\par
The linear increase in the total magnetic moment with increasing Fe content is reflected in the magnetic transition temperatures shown in Fig.~\ref{fig:data} and Table~\ref{tab:data}. For ${\rm x}\leq0.125$ the transition temperature is zero and thereafter we find a monotonous increase of the transition temperature as the amount of Fe increases. The transition temperature reaches its maximum for ${\rm x}=1$, i.e. for pure FeSb$_{3}$ for which $T_{c}=70$~K. We note that the $T_{c}$ of FeSb$_{3}$ has been reported to be 175~K.\cite{Rasander} However, the higher transition temperature in Ref.~\onlinecite{Rasander} was obtained from a scalar relativistic full-potential calculation. Here we used full relativistic calculations within the ASA. It is often found that using the scalar relativistic full-potential approach yield stronger magnetic interactions and consequently higher transition temperatures compared to the full relativistic ASA approach utilised here, due to the additional spin-flip terms that mix the majority and minority spin channels in the latter case. Even so, we find that overall the magnetic interactions yield rather low transition temperatures; much lower than possible operating temperatures of Co$_{1-{\rm x}}$Fe$_{{\rm x}}$Sb$_{3}$-based thermoelectric devices. As an additional test, we also performed self-consistent total energy calculations of FeSb$_3$ in the paramagnetic state using the disordered local moment (DLM) magnetic state within CPA. The total energy of the DLM state was found to be $\approx$ 15 meV/f.u higher than the FM state. The transition temperature in the mean field approximation (MFA) has the expression $k_BT_c^{MFA}=\frac{2}{3} \Delta E$, and using the energy difference between FM and DLM state yields 113 K, consistent with the Monte Carlo values using the exchange parameters since MFA always overestimate the value of the transition temperature.

\begin{table}[t]
\caption{\label{tab:data} Calculated magnetic moments for the skutterudite cell ($m_{tot}$), magnetic moments projected on to Co ($m_{\rm Co}$) and Fe ($m_{\rm Fe}$) atoms from full relativistic ASA calculations, as well as calculated transition temperatures obtained by Monte Carlo simulations ($T_{c}$). The errors in the transition temperatures are about $\pm3$~K.}
\begin{ruledtabular}
\begin{tabular}{ccccc}
x & $m_{\rm Co}$ ($\mu_{B}$) & $m_{\rm Fe}$ ($\mu_{B}$) & $m_{tot}$ ($\mu_{B}$) & $T_{c}$ (K) \\
 \hline
0.000	&	0.00	&		-		&	      0.00	         &	0 \\
0.125	&	-0.00	&		0.00	&		0.00      &        0\\
0.250	&	-0.01	&		0.52	&		0.51	&       12\\
0.375	&	-0.01	&		0.68	&		1.06	&       22\\
0.500	&	-0.01	&		0.76	&		1.59      &      28\\
0.625	&	-0.02	&		0.83	&		2.14	&	42\\
0.750	&	-0.02	&		0.89	&		2.70	&	55\\		
0.875	&	-0.03	&		0.93	&		3.23	&	65\\
1.000	&	-	&		0.96	&		3.74	&      70\\
 \end{tabular}
\end{ruledtabular}
\end{table}

\section{Summary and Conclusions}\label{sec:conclusions}
We have performed first principles density functional calculations in order to investigate the electronic and magnetic properties of Co$_{1-{\rm x}}$Fe$_{{\rm x}}$Sb$_{3}$ for $0\leq{\rm x}\leq1$. We find that for low Fe concentrations (${\rm x}\leq0.125$) magnetism is non-existent or at least very weak. For larger Fe concentrations we find that both the magnetic moment and transition temperatures in Co$_{1-{\rm x}}$Fe$_{{\rm x}}$Sb$_{3}$ increase monotonically with increasing Fe content. The maximum transition temperature of 70~K is found for FeSb$_{3}$. The ferromagnetism in Co$_{1-{\rm x}}$Fe$_{{\rm x}}$Sb$_{3}$ is driven by the build up of states close to the Fermi-level with increasing Fe content which makes it more and more favourable to form a ferromagnetic solution. We note that the observed paramagnetic behavior observed in Co$_{1-{\rm x}}$Fe$_{{\rm x}}$Sb$_{3}$\cite{Yang2000} is found for Fe concentrations where we find no, or possibly very weak, ferromagnetism. If ferromagnetism is weak it is likely that a transition to a paramagnetic state occurs at low temperature. In the case of FeSb$_{3}$, we believe that the observed paramagnetic behavior can be explained by spin fluctuations that renormalize the magnetic state obtained within density functional theory, as suggested by Xing {\it et al.}\cite{Xing2015}

\section{Acknowledgements}
This work was financed through the VR (the Swedish Research Council) and GGS (G\"oran Gustafsson Foundation). The computations were performed on resources provided by the Swedish National Infrastructure for Computing (SNIC) at the National Supercomputer Centre in Link{\"o}ping (NSC).

\bibliography{cofesb3}
\end{document}